\documentclass[a4paper,11pt]{article}
\pdfoutput=1 

\usepackage{jcappub} 

\usepackage[T1]{fontenc} 

   \newcommand{\exclude}[1]{}
\newcommand{\be}{\begin{eqnarray}}
\newcommand{\ee}{\end{eqnarray}}

\usepackage{graphicx,amsmath,amssymb,bm}
\usepackage{psfrag}
 \usepackage{feynmp}
\usepackage{hyperref}
\usepackage{enumitem}

\providecommand{\d}{}
\renewcommand{\d}{{\rm d}}

\providecommand{\d}{}
\renewcommand{\d}{{\rm d}}

\newcommand{\beq}{\begin{equation}}
\newcommand{\eeq}{\end{equation}}

\def\ra{\rangle}
\def\la{\langle}

\newcommand{\GeV}{\text{ GeV}}

\title{ Solar Extreme UV  radiation  and quark nugget dark matter model} 
\author{Ariel  Zhitnitsky}
\affiliation{ Department of Physics \& Astronomy, University of British Columbia, 
Vancouver, B.C. V6T 1Z1, Canada} 
 \abstract{
We advocate the idea that the surprising  emission of extreme ultra violet (EUV) radiation and  soft x-rays from the Sun are powered  externally by incident dark matter (DM) particles.   The energy and the spectral shape of this otherwise unexpected solar irradiation is   estimated within the quark nugget dark matter  model. This    model was originally invented as a natural explanation of  the 
observed ratio $\Omega_{\rm dark} \sim   \Omega_{\rm visible}$ when the DM and visible matter densities assume the same order of magnitude  values. 
 This generic consequence of  the model  is a result 
of the common   origin of  both types of matter   which are formed     during  the same  QCD transition and both proportional to the same fundamental dimensional parameter $\Lambda_{\rm QCD}$.

We also present arguments suggesting that the transient brightening-like   ``nanoflares" in the Sun may be related to the annihilation events which inevitably occur in the solar atmosphere within this dark matter scenario. 
}

\begin{document}
\maketitle
\flushbottom
\section{Introduction}

\begin{figure*}
	\centering
	 		\includegraphics[width=1.0\textwidth]{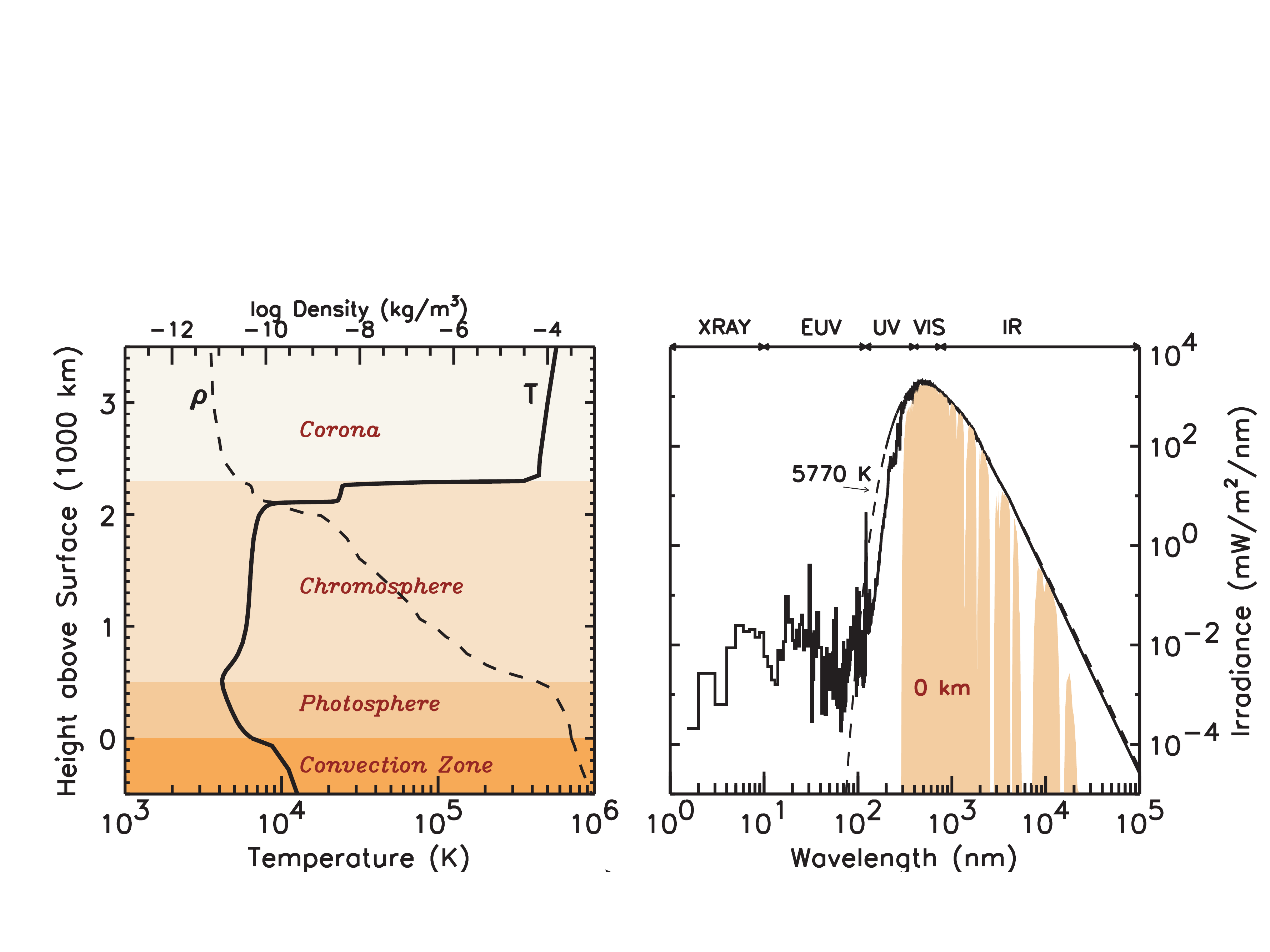}
		\label{observations}
	 	\caption{Left: The temperature distribution of the inner and outer Sun. The drastic changes occur in vicinity of 2000km.
Right:
the unexpected deviation from the thermal distribution in the extreme ultraviolet (EUV) and soft x -rays in the solar spectrum constitutes the celebrated solar corona problem. This EUV and x-ray radiation is originated from chromosphere, transition and corona regions. The total EUV intensity  represents a  small $\sim (10^{-7}-10^{-6})$ portion of the solar irradiance. The plots are taken from \cite{LZ:2003}.}
\end{figure*}

A variety of anomalous solar phenomena still defy conventional theoretical understanding. For example, the detailed physical processes that heat the outer atmosphere of the Sun to $10^6$K remain  a major open issue in astrophysics \cite{testa:2015}. In  the past, numerous theoretical models have been persuaded, but they failed to provide a quantitative explanation. We mention a few ideas related, e.g., to neutrons, interplanetary matter, acoustic waves and gravity waves 
\cite{testa:2015,Nature:1970,SolPhys:1969,ApJ:1963}. 
Therefore, after several decades of research, it may be that the answer lies in new physics. A first suggestion was made in 2002  \cite{LZ:2003}, assuming gravitationally trapped of radiatively decaying massive axion-like particles being produced in the hot solar core. 
The main goal  of the present work is to formulate a new proposal 
which may  potentially   resolve  the old  renowned  puzzle (since 1939)   known in the community under the name ``the Solar Corona Mystery". 
  This persisting puzzle is characterized by the following observed anomalous behaviour of the sun \cite{LZ:2003},\cite{KZ:2008}:
\begin{itemize} 
\item The quiet Sun emits an extreme ultra violet (EUV) radiation  with a photon energy of order of hundreds of {\rm eV  } which cannot be explained in terms of  any conventional astrophysical phenomena;  
\item The total energy output of the corona is about $(10^{-7}-10^{-6})$ times that of the photosphere ($L_{\rm corona} \sim 10^{27}$~erg/s ), which never drops to zero as time evolves. So far, no viable conventional mechanism(s) could explain this far beyond thermal equilibrium emission of  radiation; 
\item Spatially, the  unexpected EUV emission occurs near the transition region (TR), about 2000 km above the solar surface, see Fig. \ref{observations}. At first sight, it could be interpreted as the emission from a hot gas with the temperature $T\simeq 3 \cdot 10^5$~K. Though, how the underlying gas heating takes place is a long lasting question for solar physics.   The TR is the most spectacular place in the Sun, since it is where the mysterious temperature inversion appears;
\item At the transition region, the (quiet Sun) temperature continues to rise very steeply  until it reaches  a few $10^{6}$~K, i.e., being a few 100 times hotter above the underlying  photosphere, and this within an atmospheric layer thickness of only 100 km or even much less, see Fig. \ref{observations}. 
One should keep in mind that the density  at the place where  this step occurs is about $10^{-(12\pm 1)}$~g/cm$^3$, i.e. an excellent vacuum, which actually does not facilitate a conventional explanation of this observation.
  \end{itemize} 
Following conventional physics, it is fair  to say that {\it everything above the photosphere is not supposed to be there at all}. Therefore, the corona / chromosphere heating problem cannot be solved without invoking energization processes.

\exclude{
To reiterate the same claim. The main puzzles with the solar corona and the very thin transition
region between the corona and the underlying chromosphere (the least understood  region of the solar atmosphere) are the following:
The hot corona cannot be in equilibrium with the $\sim$300 times cooler solar surface underneath. In order to maintain the quiet Sun high temperature corona some non-thermally supplied energy must be dissipated in the upper atmosphere. 
One must explain the huge temperature gradient in the chromosphere $\leftrightarrow$ transition region  $\leftrightarrow$ corona, which itself is highly counterintuitive  if one attempts to explain such strange behaviour in terms of the conventional physics, using known and  well-established technical  tools.  
}

Qualitatively, this peculiar behaviour of the Sun atmosphere is suggesting  for some external irradiation (pressure) acting continuously on the whole Sun. Interestingly, also transient heating events of active regions (AR), including the concepts of ``micro-events", ``micro-flares" and ``nanoflares" \cite{Parker}, have been previously considered to be of potential interest for understanding the coronal heating mechanism \cite{warren:2010} because  they may give rise to a basal background heating near the solar surface. 

In the present work we advocate a drastically different scenario when  the energy deposition is originated from outside the system, in contrast with previously  considered proposals when  the energy is originated from the   deep  dense  regions of the sun. 
We want to argue that  the observed  peculiar behaviour  might be   intimately  related to this  fundamentally  distinct  scenario  when the extra source of the energy is associated with 
  the dark matter nuggets continuously entering the sun from outer space. 
A large amount of   energy is available  in our proposal as  result  of  huge energy deposition of such dark matter constituents (represented  in the model \cite{Zhitnitsky:2002qa} by  the quark nuggets made of matter and {\it antimatter})  before being disintegrated as described  below. 

In fact,  our proposal is largely motivated by the following numerical observation. 
It has been known  for quite some time   that the  total intensity of the  observed EUV and soft x-ray radiation (averaged over time)   can be estimated as follows,
\be
\label{estimate}
   L_{\odot ~  (\rm from ~Corona)}  \sim 10^{30}\cdot \frac{\rm GeV}{\rm  s} \sim 10^{27}  \cdot  \frac{\rm erg}{\rm  s}.
 \ee
 The corresponding flux from solar corona measured on the Earth is estimated as 
  \be
  \label{estimate1}
  \Phi_{\oplus \rm   (from ~Corona)} \sim 10^3\cdot \frac{\rm GeV}{\rm cm^2\cdot s}\sim 10^{-7}\cdot \frac{\rm W}{\rm cm^2},  
    \ee
  which represents about $(10^{-7}-10^{-6})$ fraction of the  solar luminosity. 
    It turns out that if one estimates   the extra energy being produced within the dark matter scenario  (originally suggested in \cite{Zhitnitsky:2002qa}
    for completely different purposes)  one obtains  the total extra energy $\sim 10^{27}{\rm erg}/{\rm  s}$    which 
precisely reproduces  (\ref{estimate})   for  the   observed EUV and soft x-ray intensities. One should add that the estimate   $\sim 10^{27}{\rm erg}/{\rm  s}$  for extra energy  is derived  exclusively in terms of known  dark matter density $\rho_{\rm DM} \sim 0.3~ {\rm GeV cm^{-3}}$ and dark matter  velocity $v_{\rm DM}\sim 10^{-3}c $ surrounding the sun  
 without adjusting any  parameters of the model, see  section \ref{AQN} below with relevant estimates.  We  interpret this ``numerical coincidence"  as an additional hint supporting our,   naively   ``very speculative" (but in our view  very sound) proposal.

We shall argue below  that the unusual features of the solar EUV and   soft x-rays, being emitted continuously and / or burst-like,  can be naturally explained by  the  quark nugget dark matter model
which was originally formulated   in \cite{Zhitnitsky:2002qa} for  completely different purposes.    To be more specific, this model was invented   as a natural explanation of  the observed ratio $ \Omega_{\rm dark}\sim \Omega_{\rm visible}$.  The similarity between  dark matter $ \Omega_{\rm dark}$ and the visible matter $\Omega_{\rm visible}$  densities   strongly suggests that both types of matter  have been formed  during the same cosmological epoch, which must be the QCD transition as the baryon mass $m_p$ which represents the visible portion of the matter $\Omega_{\rm visible}$ is obviously  proportional to $ \Lambda_{\rm QCD}$, while the contribution related to the Higgs   physics represents only a
minor contribution to the proton mass.  

Before turning to the details of the EUV emission from the sun within this framework  we will first give a 
brief overview of the quark nugget dark matter model in Section \ref{sec:QNDM}. In Section 
\ref{AQN} we argue that the total energy deposition  within this model is of order $(10^{-7}-10^{-6})$ times that of the photosphere,
which is precisely the observed ratio (\ref{estimate}) as mentioned above. Furthermore, we also estimate  the   spectrum of the radiation due to this mechanism. We find   it is  consistent with observations corresponding 
to the plasma temperature $T\simeq 10^6$K. In Section \ref{nanoflares} we make one step further and present  few arguments  suggesting  that the observed ``nanoflares" in quiet regions can be identified 
with annihilation events of the anti-nuggets in our framework as they fit  features observed by SoHO/EIT instrument\footnote{We want to avoid any confusions with the terminology and refer to ``nanoflares"  as the ``micro-events" in {\it quiet} regions of corona to be contrasted with ``micro-flares" which are significantly larger in scale and observed almost exclusively  in {\it active} regions, see section  \ref{nanoflares} for the precise  definitions adopted in the present  work.}. 
In  Section \ref{sun} we explain the drastic differences between the collisions of the nuggets with Earth and the Sun, suggesting that the Sun is  an ideal system to test this dark matter model. In concluding section  \ref{speculations} we
  make  few speculative comments  on possible relations of our framework with other mysterious and puzzling phenomena which are  observed,  but    not  understood yet.

\section{Axion Quark Nugget (AQN) dark matter model}
\label{sec:QNDM}
The idea that the dark matter may take the form of composite objects of 
standard model quarks in a novel phase goes back to quark nuggets  \cite{Witten:1984rs}, strangelets \cite{Farhi:1984qu}, nuclearities \cite{DeRujula:1984axn},  see also review \cite{Madsen:1998uh} with large number of references on the original results. 
In the models \cite{Witten:1984rs,Farhi:1984qu,DeRujula:1984axn,Madsen:1998uh}  the presence of strange quark stabilizes the quark matter at sufficiently 
high densities allowing strangelets being formed in the early universe to remain stable 
over cosmological timescales.    There were a number of problems with the original idea\footnote{\label{first-order}In particular, the first order phase transition is a required feature of the system for the strangelet  to be formed during the QCD phase transition.  However it is known by now that the QCD transition is a crossover rather than the first order phase transition as the recent lattice results \cite{Aoki:2006we} unambiguously show. Furthermore, the strangelets 
will likely evaporate on the Hubble time-scale even if they had been formed \cite{Alcock:1985}.} and we refer to the review paper \cite{Madsen:1998uh} for the details. 

The AQN model in the title of this section stands for the axion quark nugget   model \cite{Zhitnitsky:2002qa}   to emphasize on essential role of the axion field in the construction and to avoid confusion with earlier models    \cite{Witten:1984rs,Farhi:1984qu,DeRujula:1984axn,Madsen:1998uh} mentioned above.
The  AQN  model suggested in \cite{Zhitnitsky:2002qa} is drastically different from previous proposals in two key aspects:\\
1. There is an  additional stabilization factor in the AQN  model provided    by the axion domain walls
  which are copiously produced during the QCD transition\footnote{\label{ext_pressure}The direct consequence of this additional element  is that  the first order phase transition is not required for the nuggets to be formed as the axion domain wall plays the role of the squeezer. Furthermore, the argument  related to the fast evaporation of the strangelets as mentioned in  footnote \ref{first-order}   is not applicable for the  AQN model \cite{Zhitnitsky:2002qa} because the  vacuum ground state energies inside and outside the nuggets are drastically different. Therefore these two systems can coexist only in the presence of the additional external pressure provided by the axion domain wall.}.\\
  2. The AQN  could be 
made of matter as well as {\it antimatter} in this framework as a result of separation of charges, see original papers \cite{Zhitnitsky:2002qa, Oaknin:2003uv,Zhitnitsky:2006vt} and short review 
\cite{Lawson:2013bya}. 

 The most  important astrophysical implication  of these new aspects   relevant for the present studies of the EUV radiation 
    is that quark nuggets made of  antimatter
 store a huge amount of energy which can be released when the anti-nuggets hit the solar surface and get annihilated.  This feature 
 of the AQN model is unique and is not shared by any other dark matter models because the dark matter in AQN model is made  of the same quarks and antiquarks of the standard model (SM). One should also remark here that the   annihilation events of the anti-nuggets with visible matter  may  produce a number of other observable effects in different  circumstances such as  rare events of annihilation of anti-nuggets with  visible matter     in the centre of galaxy, or in the    Earth atmosphere,  see some comments and references   at the end of this   section.

The basic idea of  the  proposal \cite{Zhitnitsky:2002qa} can be explained   as follows: 
It is commonly  assumed that the universe 
began in a symmetric state with zero global baryonic charge 
and later (through some baryon number violating process, the so-called baryogenesis) 
evolved into a state with a net positive baryon number. As an 
alternative to this scenario we advocate a model in which 
``baryogenesis'' is actually a charge separation process 
in which the global baryon number of the universe remains 
zero. In this model the unobserved antibaryons come to comprise 
the dark matter in the form of dense nuggets of quarks and antiquarks in colour superconducting (CS) phase.  
  The formation of the  nuggets made of 
matter and antimatter occurs through the dynamics of shrinking axion domain walls, see
original papers \cite{Liang:2016tqc,Ge:2017ttc} with the details. 

 The  nuggets, after they formed,  can be viewed as the  strongly interacting and macroscopically large objects with a  typical  nuclear density 
and with a typical size $R\sim (10^{-5}-10^{-4})$ cm determined by the axion mass $m_a$ as these two parameters are linked, $R\sim m_a^{-1}$.
This  relation between the size of nugget $R$ and the axion mass $m_a$  is a result of the equilibration between the axion domain wall pressure and the Fermi pressure 
of  the dense quark matter  in CS phase. 
One can easily    estimate a typical  baryon charge $B$ of  such  macroscopically large objects as the typical density of matter in   CS phase  
is only few times the   nuclear density. Therefore, a typical baryon charge of a nugget
ranges from $B\sim 10^{24}$ to $B\sim 10^{28}$, depending on the mass of the axion, yet to be discovered.  The corresponding mass $M$ of the nuggets  can be estimated as $M\sim m_pB$, where $m_p$ is the proton mass.  

 This  model is perfectly consistent with all known astrophysical, cosmological, satellite and ground based constraints within the parametrical range mentioned   above. It is also consistent with known constraints from the axion search experiments. Furthermore, there is a number of frequency bands where some excess of emission was observed, but not explained by conventional astrophysical sources. Our comment here is that this model may explain some portion, or even entire excess of the observed radiation in these frequency bands, see short review \cite{Lawson:2013bya} and additional references at the end of this section.

Another key element of this model is   the coherent axion field $\theta$ which is assumed to be non-zero during the QCD transition in early Universe.  If the fundamental $\theta$ parameter of QCD were identically zero  
during the formation time, equal numbers of nuggets 
made of matter and antimatter would be formed.  However, the fundamental $\cal CP$ violating processes associated 
with the $\theta$ term in QCD   result in the preferential formation of 
antinuggets over nuggets\footnote{This preference  is essentially 
determined by the sign of initial $\theta_0$ before it relaxes to zero due to the axion dynamics.}.  This source of strong $\cal CP$ violation is no longer available at the present epoch as a result of the axion dynamics  when $\theta$ eventually relaxes to zero, see original papers \cite{axion,KSVZ,DFSZ} and  recent reviews   
  \cite{vanBibber:2006rb, Asztalos:2006kz,Sikivie:2008,Raffelt:2006cw,Sikivie:2009fv,Rosenberg:2015kxa,Graham:2015ouw,Ringwald2016} on the axion dynamics and recent searches of the axions. 
      
      As a result of these $\cal CP$ violating processes the number of nuggets and anti-nuggets 
      being formed would be different. This difference is always of order of one effect   \cite{Liang:2016tqc,Ge:2017ttc} irrespectively to the parameters of the theory, the axion mass $m_a$ or the initial misalignment angle $\theta_0$. As a result of this disparity between nuggets and anti nuggets   a similar disparity would also emerge between visible quarks and antiquarks.  
      
This disparity implies that  the total number of  antibaryons  will be less than the number of baryons in   early universe plasma.
These anti-baryons will be soon
annihilated away leaving only the baryons whose antimatter 
counterparts are bound in the excess of antiquark nuggets and are thus 
unavailable to annihilate. All asymmetries    are of order of 
one effects. This  is precisely  the reason why the resulting visible and dark matter 
densities must be the same order of magnitude \cite{Liang:2016tqc,Ge:2017ttc}
\be
\label{Omega}
 \Omega_{\rm dark}\sim \Omega_{\rm visible}
\ee
as they are both proportional to the same fundamental $\Lambda_{\rm QCD} $ scale,  
and they both are originated at the same  QCD epoch.
  If these processes 
are not fundamentally related the two components 
$\Omega_{\rm dark}$ and $\Omega_{\rm visible}$  could easily 
exist at vastly different scales.

 Another fundamental ratio (along with 
$\Omega_{\rm dark} \sim  \Omega_{\rm visible}$  discussed above)
is the baryon to entropy ratio at present time
\be
\label{eta}
\eta\equiv\frac{n_B-n_{\bar{B}}}{n_{\gamma}}\simeq \frac{n_B}{n_{\gamma}}\sim 10^{-10}.
\ee
In our proposal (in contrast with conventional baryogenesis frameworks) this ratio 
is determined by the formation temperature $T_{\rm form}\simeq 41 $~MeV  at which the nuggets and 
antinuggets compete their formation, when all anti baryons get annihilated and only the baryons remain in the system, see details in    \cite{Liang:2016tqc}. We note that $T_{\rm form}\approx \Lambda_{\rm QCD}$. This temperature    is determined by the observed  ratio (\ref{eta}). The $T_{\rm form}$  assumes a typical QCD value, as it should as there are no any small parameters in QCD. 

One can recapitulate the same claim  in terms of the cosmological time $t$ instead of the temperature $T$. As it is known the QCD transition happens at $T\simeq 170$~MeV which corresponds to cosmic time $t_0\simeq 10^{-4}$~s. During the radiation epoch the temperature  scales as $T\sim t^{-1/2}$.
Therefore, the formation of the nuggets is completed around $t_{\rm form}\simeq 10^{-3}$~s which corresponds to the formation temperature $T_{\rm form}\simeq 41 $~MeV when the baryon to entropy ratio assumes its present value (\ref{eta}). 

Unlike conventional dark matter candidates, such as WIMPs 
(Weakly interacting Massive Particles) the dark-matter/antimatter
nuggets are strongly interacting and macroscopically large as already mentioned. 
However, they do not contradict any of the many known observational
constraints on dark matter or
antimatter  for three main reasons~\cite{Zhitnitsky:2006vt}:
\begin{itemize} 
\item They carry  very large baryon charge 
$|B|  \gtrsim 10^{24}$, and so their number density is very small $\sim B^{-1}$.  
 As a result of this unique feature, their interaction  with visible matter is highly  inefficient, and 
therefore, the nuggets are perfectly qualify  as  DM  candidates. In particular, the quark nuggets  essentially decouple 
from CMB photons, and therefore, they do not destroy conventional picture for the structure formation; 
\item The core of the  nuggets have nuclear densities. Therefore, the relevant  effective interaction
is very small $\sigma/M \sim 10^{-10}$ ~cm$^2$/g. Numerically, it is  comparable with conventional WIMPs values.
Therefore, it is consistent  with the typical astrophysical
and cosmological constraints which  are normally represented as 
$\sigma/M<1$~cm$^2$/g;
\item The quark nuggets have  very  large binding energy due to the   large    gap $\Delta \sim 100$ MeV in CS phases.  
Therefore, the baryon charge is so strongly bounded in the core of the nugget that  it  is not available to participate in big bang nucleosynthesis
(\textsc{bbn})  at $T \approx 1$~MeV, long after the nuggets had been formed. 
\end{itemize} 
\exclude{
We emphasize that the weakness of the visible-dark matter interaction 
in this model is due to a  small geometrical parameter $\sigma/M \sim B^{-1/3}$ 
  which replaces 
the conventional requirement of sufficiently weak interactions for WIMPs. 
}

  It should be noted that the galactic spectrum 
contains several excesses of diffuse emission the origin of which is unknown, the best 
known example being the strong galactic 511~keV line. If the nuggets have the  average  baryon 
number in the $\langle B\rangle \sim 10^{25}$ range they could offer a 
potential explanation for several of 
these diffuse components.  
\exclude{(including 511 keV line and accompanied   continuum of $\gamma$ rays in 100 keV and few  MeV ranges, 
as well as x-rays,  and radio frequency bands). }
It is important to emphasize that a comparison between   emissions with drastically different frequencies in such a computations 
 is possible because the rate of annihilation events (between visible matter and antimatter DM nuggets) is proportional to 
one and the same product    of the local visible and DM distributions at the annihilation site. 
The observed fluxes for different emissions thus depend through one and the same line-of-sight integral 
\be
\label{flux1}
\Phi \sim R^2\int d\Omega dl [n_{\rm visible}(l)\cdot n_{DM}(l)],
\ee
where $R\sim B^{1/3}$ is a typical size of the nugget which determines the effective cross section of interaction between DM and visible matter. As $n_{DM}\sim B^{-1}$ the effective interaction is strongly suppressed $\sim B^{-1/3}$. The parameter $\la B\ra\sim 10^{25}$  was fixed in this  proposal by assuming that this mechanism  saturates the observed  511 keV line   \cite{Oaknin:2004mn, Zhitnitsky:2006tu}, which resulted from annihilation of the electrons from visible matter and positrons from anti-nuggets.   Other emissions from different bands  are expressed in terms of the same integral (\ref{flux1}), and therefore, the  relative  intensities  are unambiguously and completely determined by internal structure of the nuggets which is described by conventional nuclear physics and basic QED.  In particular, the excess of the diffuse  gamma ray emission   in  $1-20$ MeV range observed by COMPTEL  
might be related with these annihilation processes of the galactic visible electrons with anti-nuggets, as argued in \cite{Lawson:2007kp,Forbes:2009wg}. For further details see the original works 
 \cite{Oaknin:2004mn, Zhitnitsky:2006tu,Forbes:2006ba, Lawson:2007kp,
Forbes:2008uf,Forbes:2009wg,Lawson:2012zu} and a short overview  \cite{Lawson:2013bya} with specific computations of diffuse galactic radiation  in different frequency bands.

The studies which are most relevant for our  present work  is the analysis  \cite{Gorham:2015rfa,Lawson:2015cla} on neutrino emission from the Sun as a result of
    interaction of the antinuggets with  solar environment.   The basic claim of ref.  \cite{Gorham:2015rfa} is as follows: if one assumes that
the neutrino spectrum (as a result of annihilation of   anti nuggets with matter in Sun)  is similar to the spectrum observed in studies of low -energy $p\bar{p}$ 
annihilation, then anti-quark nuggets   cannot account for more than $1/5$ of the dark matter flux. 

This claim has been dismissed in ref. \cite{Lawson:2015cla}
by emphasizing that anti-nuggets cannot be treated as an usual antimatter in conventional hadronic phase as the quarks in nuggets (and   antiquarks in anti nuggets) belong to   CS phase rather than to the  hadronic phase we are familiar with. In particular, the Nambu-Goldstone bosons in CS phase are 5-10 times lighter than their counterparts (such as conventional $\pi$ mesons) in the hadronic phase. This observation  leads to the profound consequences as  the neutrino  spectrum is expected to be in  the 10 MeV range, in 
contrast with the 50 MeV scale associated with conventional hadronic decays as 
used in the analysis of \cite{Gorham:2015rfa}. Therefore, the quark nugget dark matter model is  fully consistent with present neutrino flux measurements 
as stringent constraints from SuperK are not sensitive to the low energy neutrinos in the 10 MeV range \cite{Lawson:2015cla}.

  \section{Anti-nuggets  in  the Sun  }\label{AQN} 
  We first present few basic relations from refs \cite{Gorham:2015rfa,Lawson:2015cla} on the energy budget due to the capture of the anti-nuggets by the Sun.
  After that we focus on the spectral properties of the radiation from the anti nuggets when they interact with the matter from the Sun.
  
  The impact parameter for capture of the nuggets by the Sun can be estimated as
  \be
  \label{capture}
  b_{\rm cap}\simeq R_{\odot}\sqrt{1+\gamma_{\odot}}, ~~~~ \gamma_{\odot}\equiv \frac{2GM_{\odot}}{R_{\odot}v^2},
  \ee
  where $v\simeq 10^{-3}c$ is a typical velocity of the nuggets. One can easily see that  $\gamma_{\odot}\gg 1$
  which   should be contrasted with the corresponding parameter on Earth, $\gamma_{\oplus} \ll 1$.
   Nuggets in the solar atmosphere will  be decreasing their mass as result of annihilation, decreasing  their kinetic energy and velocity as result of ionization and radiation. 
  
    Assuming that $\rho_{\rm DM} \sim 0.3~ {\rm GeV cm^{-3}}$ and using the capture impact parameter (\ref{capture}), one can estimate 
  the total energy flux due to the complete annihilation of the nuggets,
   
  \be
  \label{total_power}
   L_{\odot ~  \rm (AQN)}\sim 4\pi b^2_{\rm cap}\cdot v\cdot \rho_{\rm DM}   
  \simeq 3\cdot 10^{30} \cdot \frac{\rm GeV}{\rm  s}\simeq 4.8 \cdot 10^{27} \cdot  \frac{\rm erg}{\rm  s}, 
  \ee
   where we substitute  constant $v\simeq 10^{-3}c$  to simplify numerical  analysis. 
   This estimate is very suggestive as it roughly coincides with the total EUV  energy output (\ref{estimate}) from corona which is hard to explain in terms of conventional astrophysical sources as highlighted in the Introduction. Precisely this ``accidental  numerical coincidence" was the main motivation   to put forward the idea that  (\ref{total_power}) represents a new source of energy feeding the EUV radiation as advocated in this work. 
   This source of the energy is unique to the AQN dark matter model as the energy is released as result of the annihilation of the antiquarks from anti nuggets with conventional baryonic matter in the solar Corona. 
   
   The corresponding flux  due to the anti-nugget annihilations measured on the Earth
   is estimated as 
  \be
  \label{total_power1}
  \Phi_{\oplus~  \rm (AQN)}\sim 10^3\frac{\rm GeV}{\rm cm^2\cdot s}\sim 1.6\cdot 10^{-7}\frac{\rm W}{\rm cm^2},
  \ee
  which should be compared with observed flux (\ref{estimate1}). 
    The crucial observation of the present work is that  while  the total    energy due to the annihilation of the anti-nuggets 
   is indeed very small, nevertheless the anti-nuggets  produce EUV spectrum characterized by the temperature $T\sim 10^6$K as explained below.
   Such spectrum observed in Corona is  hard to explain by any conventional astrophysical processes as
   stated in the Introduction. 
   
   One should emphasize that the estimates (\ref{total_power}) and (\ref{total_power1}) for the total intensity as well as 
   the estimate   for a typical temperature given in next subsection \ref{spectrum}, 
   are not sensitive to the size distribution of the nuggets. This is because  the  estimates (\ref{total_power}) and (\ref{total_power1}) represent the total energy input due to the compete nugget's annihilation, while their total baryon charge is determined by the dark matter flux  $\rho_{\rm DM} \sim 0.3~ {\rm GeV cm^{-3}}$. 
   
   In contrast, there are many observables, to be mentioned in next sections \ref{interpretation} and  \ref{nanoflares} which are highly sensitive to the size distribution of the  nuggets because the corresponding observables are  expressed in terms of the   baryon charge $B$ of a nugget or   
   its size $R\sim B^{1/3}$. 
   
   \subsection{\label{spectrum} The Basic Features of the Spectrum} 
   The goal here is to estimate the typical spectral properties  of the radiation due to the annihilation of the anti-nuggets in solar environment. 
   We follow ref. \cite{Forbes:2008uf} in our estimation of the temperature of the 
anti-nuggets as they just enter the solar Corona.

The basic idea is to equate the total surface emissivity $ F_{\text{tot}}$ with 
the rate at which annihilations deposit energy $F_{\text{ann}}$
within a given environment for each given anti-nugget. The total surface
emissivity as well as the spectral density have been computed in \cite{Forbes:2008uf}. It is mostly determined 
by photon emission from the electrosphere of the nugget, and it is given by
\begin{equation}
\label{eq:P_t}
F_{\text{tot}} =  \frac{\d{E}}{\d{t}\;\d{A}} \simeq  
\frac{16}{3} \frac{T^4\alpha^{5/2}}{\pi}\sqrt[4]{\frac{T}{m}}.
\end{equation}
We should note that the spectrum of the emission is also known, and it is approximately flat in the extended region of frequencies with $\omega\leq T$. This should be  contrasted with black body radiation, see  \cite{Forbes:2008uf} for the details.

In order to maintain the overall energy balance between the emission rate  given by $F_{\text{tot}}$ and annihilation rate given by  $F_{\text{ann}}$, the nuggets must emit energy at the same rate that it is deposited through proton annihilation, $F_{\text{tot}} \simeq F_{\text{ann}}$.
 
  Note that both the rate of emission and the rate of
annihilation are expressed per unit surface area, so that the equilibrium
condition is independent of the nuggets' size, and therefore of their average 
baryon number $\langle B\rangle $. 

The rate of annihilation $F_{\text{ann}}$ is
\begin{equation}
\label{eq:Fann1}
F_{\text{ann}} = 2\GeV\cdot f (T,l)\cdot \eta (T,l) \cdot v (l) \cdot n_{\rm sun} (l)
\end{equation}
where $2\GeV \simeq  2\,m_{p}$ is the energy liberated by proton annihilation,
$v(l)$ is the average speed  of the nugget at the altitude $l$ above the surface,
$n_{\rm sun}(l)$ is the average nucleon density at the altitude $l$, $f(T)\leq 1$
is a suppression  factor  due to the possibility of reflection from
the sharp quark-matter surface determined by the internal nugget's temperature $T$. Finally, $\eta(T, l)\geq 1$ is an enhancement factor in the annihilation rate due to ionization of the surrounding plasma 
and very large charge of the anti-nugget $Q$ which is build in during the motion of the anti-nuggets  in the solar corona.
We estimate the corresponding charge $Q$ in Appendix \ref{ionization}. 

Now we want to present some simple numerical   estimates 
for the altitude  $l\sim 3000$ km above the solar surface,   where the density is sufficiently  small and plasma effects cannot drastically change our estimate below. For order of magnitude estimates we  take $n_{\rm sun}(l)\sim 10^{10} {\rm cm}^{-3}$,
$v\sim 10^{-3} c$. Suppression factor $f\simeq 0.1$ was estimated in  \cite{Forbes:2008uf}  for low temperature and low density  environment,  while enhancement factor $\eta(T)$ was not even considered previously in  \cite{Forbes:2008uf}  as anti-nuggets were assumed to be neutral. It is very hard to estimate factor $\eta(T)$
as it depends on many properties of the charged anti-nuggets and surrounding magnetized plasma. We  presented few  arguments  in Appendix  \ref{ionization} that the corresponding enhancement factor  $\eta(T)$ due to the building of a large internal charge of the nugget could be very significant, especially in denser regions. 
For simplicity of this order of magnitude estimate we assume that 
these two factors approximately neutralize each other for the low density plasma for $n_{\rm sun}(l)\sim 10^{10} {\rm cm}^{-3}$, i.e. we assume  in eq. (\ref{eq:Fann2}) that $[f(T)\cdot \eta(T)]\simeq 1$. 
Combining these numerical values, we obtain
\begin{equation}
\label{eq:Fann2}
F_{\text{ann}} \sim \frac{10^{17}~\text{GeV}}{\text{cm}^2\cdot\text{s}}
\cdot \left(\frac{f\cdot \eta}{1}\right) \cdot \left(\frac{v}{10^{-3}c}\right)\cdot
\left(\frac{n_{\rm sun}(l)}{10^{10}/\text{cm}^{3}}\right)
\end{equation}
which must be compared with the total surface emissivity~(\ref{eq:P_t}) 
which  can be represented as follows
\begin{equation}
\label{eq:Ftot}
F_{\text{tot}} \sim 10^{17}
\frac{\text{GeV}}{\text{cm}^2\cdot\text{s}}
\left(\frac{T}{ 
10^6~ \text{K}}\right)^{4+1/4}, ~~~ T\sim 10^6~K.
\end{equation}
Taking the typical values for these parameters mentioned above 
and comparing (\ref{eq:Fann2}) with (\ref{eq:Ftot}) we arrive at an estimate of   
an intrinsic nugget  temperature of $T\simeq 10^6 ~\text{K}$ for anti-nuggets travelling in the solar corona at an altitude   of $l\sim 3000$ km. 

One should emphasize that the temperature $T\simeq 10^6 ~\text{K}$ which enters   formulae (\ref{eq:Ftot}) is the internal temperature 
of the anti-nuggets, and  should not be confused with  the surrounding  plasma of the solar corona.   The thermodynamical processes  will  eventually  equilibrate the internal temperature (\ref{eq:Ftot}) with the temperature in surrounding  plasma. However, this is not an instantaneous process, and there are many mechanisms in plasma which can transfer the energy  (\ref{eq:Ftot}) to the surrounding particles. We shall not discuss the corresponding  equilibration mechanisms, as they are  well beyond the scope of the present work. To simplify our  analysis we just assume that the entire energy produced by the annihilating anti-nuggets will be  eventually transferred to the surrounding plasma.

When the nuggets are slowly descending and 
the plasma density $n_{\rm sun}(l)$ is slowly increasing, the  number of annihilation events (and corresponding energy deposit)  is also increasing.  It is expected that  the rate of energy transfer from nuggets to the surrounding plasma will be  also   increasing\footnote{\label{T}The  energy transfer from nuggets to    surrounding plasma   in this case is not expressed  by  a simple expression (\ref{eq:P_t}) describing the photon emission by  the nugget's electrosphere  at temperature $T$.  The dominant  mechanisms are expected to be much more complicated, and may include such   processes   as elastic and inelastic scatterings of the nugget's degrees of freedom with protons and electrons from plasma,  direct exchange processes, etc. } as time evolves. Precise estimations of all these effects which eventually should explain a complicated behaviour of the temperature $T(l)$ as a function of altitude $l$ are well beyond the scope of the present work and shall   not be  elaborated further.

The significance of this entire picture  is that  the anti-nuggets will be transferring their   energy due to the annihilating processes to surrounding plasma with the spectrum and intensity which are consistent with the
``mysterious puzzling features"  listed  in the Introduction. This is because  the total intensity of the EUV emission is exclusively determined by the number of anti-nuggets captured by   the Sun. It   is estimated   that the anti-nuggets will  deposit their    energy  on  the level  of $10^{-6}$ of the total solar luminosity according to (\ref{total_power}) and (\ref{total_power1}), in agreement with the observations.

\subsection{The annihilation pattern. The big picture. }\label{interpretation}

The main assumption of the present work is that a finite portion of annihilation events have occurred before the anti-nuggets entered the dense regions of the Sun\footnote{We note that  the complications related to a precise  annihilation pattern (including the ionization of the nuggets and consequently, the  complications related to the  dynamics of the nuggets  in the magnetic field)  were absolutely irrelevant for analysis in refs.\cite{Gorham:2015rfa,Lawson:2015cla} because the neutrinos will leave the system irrespectively of the  precise altitude where annihilation event occurs. This is not the case with EUV radiation which can only leave the system if annihilation occurs above (or slightly below) the photosphere. This is because the EUV photons will be quickly thermalized if annihilation occurs in dense regions below the sun surface.}.
Just these annihilation events, according to our  assumption, supply the energy source of the observed EUV radiation from the corona and the choromosphere. 

The basic picture of  the annihilation pattern can be viewed  as follows. 
The capture parameter $\gamma_{\odot}$  for the Sun is large. 
This unambiguously implies that a significant  portion of the nuggets which are not on the head on collision trajectories with the Sun,  nevertheless will be  captured by the Sun according to  (\ref{capture}). 
It also implies that a typical  length of the nugget's trajectory (where annihilation occurs) is quite long as it is   of order of $ R_{\odot}\sim 10^6$ km as argued in 
Section \ref{sun}, in huge contrast  with collision of the AQN with  Earth when $\gamma_{\oplus} \ll 1$.    In the case with collision with the Earth 
a typical length  of the AQN's    trajectory  (where annihilation effectively occurs)  would be of order of  height of the Earth's atmosphere $\sim 10^2$ km. 

As a result of this difference,  the dominant portion  of the energy due to the annihilation of the AQN  in the Earth's case will be deposited in the  deep Earth's underground  regions while the main portion of the   energy in the solar's case will be mostly deposited  in  the  corona/ chromosphere/ transition   regions   allowing  the direct observations in form of the EUV and soft x-ray emissions from that regions. There are few additional arguments (such as high ionization of the solar corona) supporting this basic assumption. We elaborate on these issues in Section \ref{sun} where we argue  that the Sun is an ideal place to study the AQN dark matter.

 To recapitulate the main assumption of this work: we assume that  the large portion of the AQN  will   loose their  masses (their initial baryon charges) and their initial kinetic energies  in Corona and Chromosphere   before entering the  dense photosphere region.  At  this  point they finally   get annihilated completely and cease to exist. 

We estimate the corresponding time scale $\tau$ where this most important portion of the annihilation occurs as follows.
\be
\Delta B\sim n_{\rm sun}(l) \cdot \left(\pi R_{\rm eff}^2\right)\cdot v  \tau\sim B,
\label{time}
\ee
where $R_{\rm eff}(l)$ is the effective size of an anti-nugget during the last moments of its existence. It does   not coincide with the nugget's size $R$ because the nuggets are electrically charged objects propagating in the ionized plasma, and the effective cross section for annihilation  is much larger than $R$, see section \ref{ionization} for the relevant estimates.   
Essentially this effective parameter accounts for complicated plasma effects (such as shock waves, bluster waves, etc) which could develop in unfriendly environment in vicinity of the moving nugget when a large number of particles from surrounding areas will be affected by the presence of the nugget, and will be involved in its dynamics. Eventually, these surrounding particles  will have a chance for successful annihilation with anti-nugget, which is effectively described by a parameter $R_{\rm eff}(l)$ in eq. (\ref{time}). 


For very rough order of magnitude numerical estimates 
we assume  $n_{\rm sun}\sim (10^{16}-10^{18}){\rm cm^{-3}}$ as a typical density of  the region where anti-nuggets effectively complete their annihilation processes and cease to exist.  For numerical estimates we assume  $R_{\rm eff}\sim 0.1 ~{\rm cm}$ at this density as estimated in Appendix \ref{ionization}. With  these effective parameters in hands 
we arrive to the following numerical estimates 
for a typical time $\tau$ during which the  anti nugget will loose  a large portion of  its  baryon charge  before entering the deep interior regions of the Sun,
\be
\frac{\Delta B}{B}\sim\big(\frac{10^{25}}{B}\big)^{\frac{1}{3}}\times \big(\frac{n_{\rm sun}(l)}{10^{17}\rm cm^{-3}} \big)\times \big( \frac{R_{\rm eff}(l)}{10^{-1}{\rm cm}} \big)^2   \times \big(\frac{v}{10^{-3}c}\big)\times   \big(\frac{\tau}{10^2\rm{s}}\big)\sim 1. 
\label{time1}
\ee
 Numerical estimate (\ref{time1}) suggests that a typical time scale when a  large portion of the annihilation occurs is of the order of $10^2$ seconds.
 This estimate, of course, strongly depends on many numerical parameters used in  (\ref{time1}) such as plasma density $n_{\rm sun}(l)$, effective radius $R_{\rm eff}(l)$,  and the initial baryon charge  $B$ of an anti-nugget.

 It is quite possible that the unusual features of the 
  the transition region (such as very fast rise of the temperature within $10-100$ km)  are intimately related to these annihilation events. It is also possible that at the  altitude of $l\sim 2000$ km      the nuggets   are  loosing their kinetic energy, their baryonic charge, their mass very efficiently when they approach the photosphere from the higher elevations. It is also  very likely that the anti-nuggets may loose their entire baryon charge  $B$ before   entering  the   dense regions  of the photosphere, which is precisely our main assumption formulated above and qualitatively supported by the estimate (\ref{time1}). 
  
  It is not our goal to speculate on these and  other  related questions, which are obviously beyond the scope of the  present work. The only comment we would like to make here      is that the estimate (\ref{time1}) on the annihilation pattern is quite sensitive to a specific size of the nugget,
   in contrast with our previous estimates (\ref{total_power}), (\ref{total_power1}) and (\ref{eq:Ftot}) which are very generic features of the framework rather than specific properties  of a model.  In fact, we present the  model-dependent estimate     (\ref{time1}) in this subsection exclusively with a  purpose  to identify the     large events of annihilation in the corona/chromosphere  (when $B$ could be as large as $B\gtrsim 3\cdot 10^{26}$) with the observed and previously analyzed  nanoflares as advocated  in section \ref{nanoflares}.

 \section{Observation  of nanoflares as evidence for    anti-nuggets in Corona  }\label{nanoflares}
 The main claim of this work is that  the corona and the chromosphere might be  powered by  the annihilation of the dark matter anti-nuggets.
  All our arguments supporting this claim,  up to this section,     were based on the consideration of the  total energy budget estimates (\ref{total_power}) and the typical spectral properties (\ref{eq:Ftot}).  Both these observables  are not sensitive to a specific size distribution of the nuggets, as it has been  already mentioned previously. 

 In this section we want to make one step further and identify the annihilation events of the anti-nuggets with the previously studied ``nanoflares", which belong to the burst-like solar activity.     Therefore, this section is much more speculative, in comparison with previous discussions  of section \ref{AQN},  as it includes an  additional assumption in form of identification 
 of the AQN annihilation events with observed ``nanoflares". 
 
      However, before we elaborate on this possible connection between the AQN annihilation events and   nanoflares we want to make a short historical detour in section \ref{nanoflares-history} on the nanoflares and their role in physics of corona as studied in previous works. In section \ref{nanoflares-nuggets} we present few arguments (based on various observations) supporting our interpretation  that  sufficiently energetic  annihilation events of the anti-nuggets with $B\gtrsim 3\cdot 10^{26}$ are behind the observed nanoflares.  Smaller   nanoflares  are  not directly observed due to  the instrumental threshold. In section \ref{nanoflares-development} we make few comments on possible future   development. We also overview in  section \ref{nanoflares-development} 
      some recent results related to the modelling of the plasma  heating   in corona when  the nanoflares    play the  key role in analysis.

 \subsection{Few historical remarks on nanoflares}\label{nanoflares-history}
 The term ``nanoflare" has been introduced by Parker in 1983 \cite{Parker}. Later on this term has been used in series of papers by Benz and coauthors
  \cite{Benz-2000,Benz-2001, Kraev-2001,Benz-2002, Benz-2003}  to advocate the idea that precisely these small ``micro-events" might be  responsible for the   heating of the   quiet solar corona.  We want to list below few important features which have been discussed in previous works  \cite{Benz-2000,Benz-2001, Kraev-2001,Benz-2002, Benz-2003} before we present our original arguments suggesting that these nanoflares 
  have the same features as   annihilation events of   anti nuggets in the corona.
  
  We follow the definition suggested in \cite{Benz-2003} and refer to nanoflares as the ``micro-events" in quiet regions of the corona, to be contrasted with ``micro flares" which are significantly larger in scale and observed in active   regions.   The term ``micro-events" refers to a short enhancement of coronal emission in the energy range of about $(10^{24}-10^{28})$erg. One should emphasize that the lower limit   gives the instrumental threshold observing quiet  regions, while the upper limit refers to the smallest events observable in active regions. Below we list most important features of 
nanoflares which became possible by observing EUV iron lines with SoHO/EIT. The main features are:\\
{\bf 1} The coronal emission measure observed in EUV iron lines fluctuates locally at time scales of few minutes in a majority of pixels including even the intracell regions of the quiet corona;\\
{\bf 2} It was reported $1.1\times 10^6$ events per hour over the whole Sun for SoHO/EIT observations \cite{Benz-2001, Benz-2002};\\
 {\bf 3} The energy output  observed by EIT on the SoHO satellite is of order of  $10\%$ of the total radiative output in the same region \cite{Benz-2002};\\
{\bf 4} To reproduce the measured  radiation loss, the observed range of nano flares (having a lower limit at about $3\cdot 10^{24}$erg is  due to the instrumental threshold) needs to be extrapolated to  energies as low as  $ 10^{22}$erg  and in some models even to $ 10^{20}$erg, see table 1 in ref.\cite{Kraev-2001};\\
{\bf 5} nanoflares and micro-flares appear in different ranges of temperature and emission measure,  see  Fig.3 in ref. \cite{Benz-2003}. While  the instrumental  limits prohibit observations at intermediate temperatures, nevertheless the authors of ref. \cite{Benz-2003} argue that 
    ``the occurrence rates of nanoflares and micro-flares are so different that they cannot originate from the same population";\\
{\bf 6} Time measurements of many nanoflares demonstrate the Doppler shift with a typical velocities (250-310) km/s, see Fig.5 in ref. \cite{Benz-2000}.
The observed line width in OV  of $\pm 140$ km/s far exceeds the thermal ion velocity which is around 11 km/s   \cite{Benz-2000}.

 \subsection{Nanoflares as the annihilation events of the anti-nuggets}\label{nanoflares-nuggets} 
 It is very tempting to identify the nanoflares reviewed in previous subsection  \ref{nanoflares-history} with the events of annihilation of the anti-nuggets inside the solar corona.  Before we present our arguments supporting this identification we would like to make the following few generic  comments.  
 
 We describe the nuggets in terms of the baryon charge $B$. Annihilation of a single baryon charge produces  the energy about 2 GeV which is convenient to express in terms of the conventional units as follows,
 \be
 \label{units}
 1~ {\rm GeV}=1.6 \cdot 10^{-10} {\rm J}=1.6\cdot 10^{-3} {\rm erg}.
 \ee 
This relation suggests that the current instrumental threshold of a nanoflare characterized by the energy  $\sim 10^{24}~{\rm erg}$ corresponds to the (anti) baryon charge of the nugget $B\approx 3\cdot 10^{26}$.  Anti-nuggets with smaller $B$  are   present in the corona  but are considered as the sub-resolution events. This claim follows from item {\bf 4} above which suggests that nanoflares with sub-resolution energies  must be present in the system to reproduce the measured radiation loss. 
 
 Indeed, according to item {\bf 4} from previous subsection  \ref{nanoflares-history},  in order to reproduce the observed radiation loss the energy of nanoflares must be extrapolated from sub-resolution events with energy $3.7\cdot 10^{20}~{\rm erg}$ to the observed events  interpolating between   $(3.1\cdot 10^{24}  - 1.3\cdot 10^{26})~{\rm erg}$.  For a different extrapolation model used in \cite{Kraev-2001} the energy varies from  $(9.8\cdot 10^{21}- 6.3\cdot 10^{25})~{\rm erg}$.  This energy window corresponds to the 
 (anti) baryon charge of the nugget $ 10^{23} \leq |B|\leq  4\cdot 10^{28}$ (or $ 3\cdot 10^{24} \leq |B|\leq  2\cdot 10^{28}$  for a different extrapolation model used in \cite{Kraev-2001}). 
 
 We want to emphasize that this window of acceptable extrapolations is largely  overlapped   with     all presently available and independent constraints on such kind of dark matter masses and baryon charges $B$, see e.g. \cite{Jacobs:2014yca,Lawson:2013bya} for review\footnote{The smallest nuggets with $B\sim (10^{23}-10^{24})$ naively  contradict to the constraints cited in \cite{Jacobs:2014yca}. However, the corresponding constraints are actually derived with the assumption that   nuggets with a definite mass (smaller than 55g) saturate the dark matter density. In contrast, we assume  that the peak  of the nugget's distribution corresponds to a larger value of mass, while the small nuggets represent a tiny portion of the total dark matter density. The same comment also applies to   the larger masses excluded by Apollo data as reviewed in  \cite{Lawson:2013bya}.
  Large nuggets with $B\sim 10^{28}$ do exist, but represent  a small portion of the total dark matter density.}. To reiterate the same claim:  the allowed window for the baryon charge $B$ (and corresponding radiation energy due to annihilation of the anti-nuggets) is   perfectly   consistent with all presently available constraints. 
 
 According to item {\bf 5} above   the nanoflares are distributed very uniformly in quiet  regions, in contrast with micro-flares
 which are much more energetic and occur exclusively in active areas. 
 It  is consistent with the dark matter   interpretation as the  anti-nuggets annihilation events (identified with nanoflares)  should be present in all areas irrespectively to the activity of the Sun.  At the same time the micro- flares  are originated in the  active zones, and therefore cannot be uniformly distributed.

 The presence of the large Doppler shift  with a typical velocities (250-310) km/s as mention in  item {\bf 6} above can be  understood  within the dark matter interpretation of the nanoflares. Indeed, the typical velocities of the nuggets entering the solar corona is about $ \sim  300 ~{\rm km/s}$, which is very close to the measured  Doppler shift.
 
 The observed   iron lines as mentioned in item {\bf 1} is also easy to interpret within our identification of the nano flares with  the annihilating   anti-nuggets. Indeed, according to 
 the estimate (\ref{time1}) the   typical event with $B\sim (10^{27}-10^{28})$ corresponding  to the observed  nanoflares  may   last few minutes. This is the time scale needed for anti-nuggets to completely annihilate their baryon charge. 
 This time-scale is obviously highly sensitive to the altitude $l$ where annihilation happens and to the initial baryon charge $B$ of the anti-nuggets.
 However, a very reasonable order of magnitude estimate (\ref{time1})  once again  supports our reasoning that  the nanoflares can be identified with the  annihilation events of   dark matter anti-nuggets. 
 
 The measured energy   and number of recorded events according to items {\bf 2} and {\bf 3} above represent only small factor of the total radiation energy  in the same solar region. 
 The interpretation of this ``apparent deficiency" is also very straightforward within  our framework. Indeed, only a  small portion of the nuggets are sufficiently large to produce the events with the energies above the instrumental threshold which can be recorded. Smaller events are likely to  occur in the corona as discussed above (see the  paragraph   after eq. (\ref{units}) with estimates on the window for B), and likely to  contribute to the total solar radiative output in this energy range, but they are not recorded due to insufficient resolution of the current instruments.   
 
 \subsection{ Recent advances  and Future  development}\label{nanoflares-development}
In this subsection we make few comments on recent results  on nanoflares and 
their role in the heating mechanisms of corona. We also   comment on  possible future  development  which may support  our interpretation of the nanoflares  as the  AQN annihilation events. 

  It would be highly desirable to estimate the baryon charge distribution of the AQNs from the theoretical side, irrespectively to the solar physics. 
  In this case one could estimate the corresponding energy distribution of the nanoflare events  in the solar corona as the baryon charge of the AQN  is unambiguously    translated into the energy of the nanoflares.  This statement can be formally expressed as follows 
  \be
  \label{distribution}
   {dN} \sim B^{-\alpha}dB\sim W^{-\alpha} dW, 
  \ee
  where $dN$ is the number of the nanoflare  events per unit time with energy between $W$ and $W+dW$ which occur as a result of complete annihilation 
  of the anti-nuggets carrying the  baryon charges between  $B$ and $B+dB$.
 These two distributions are tightly linked  
    as these two entities are related to the same AQN objects when the unit  baryon charge   generates annihilation energy (\ref{units})  
    according to our  interpretation of the observed nanoflare events. Unfortunately, the corresponding theoretical estimates of the distribution 
   $dN/dB$ are very hard to carry out in spite of the fact that we are dealing with the standard model (SM) physics with no   new fundamental parameters outside  the SM physics. 
    
    Indeed, one encounters  two  major difficulties  in the path to  produce  such a theoretical  estimate. First of all, the initial axion domain wall size distribution of the closed bubbles (which eventually form the nuggets) is unknown as discussed in the original paper \cite{Liang:2016tqc}
 devoted to the formation mechanism of the AQNs. Only the total baryon charge hidden in all  nuggets and anti-nuggets  is known as 
 it is expressed in terms of the dark matter density. In practice the problem becomes even more complicated because the axion domain wall bubble evolution  is  accompanied by  the process of accumulation of the baryon charge in its bulk as discussed in \cite{Liang:2016tqc}. It is very hard  to predict the  outcome 
 of these complicated processes depending on  the size distribution.  Needless to say that even the phase diagram at $\theta\neq 0$
 when the formation of the AQN occurs is still unknown, and the QCD lattice numerical simulations run into the fundamental obstacle  as a result of the so-called sign problem when $\theta $ parameter  and chemical potential $\mu $ assume some non-zero values. 
 
 The second major barrier    to produce a  theoretical  estimate (\ref{distribution}) irrespectively to the solar physics is a lack of  understanding of the evolution of the AQN from the moment of formation  at $T\sim 1$ GeV to the present cold Universe. The problem   here is that the    nuggets evolve in very unfriendly environment in plasma when all relevant parameters $T, \mu, \theta$ 
 were nonzero and some (smaller) nuggets may not survive the evolution, while other (larger) nuggets may drastically change their initial baryon charge   as a result of  annihilation and accretion processes. All these effects, are obviously order of one effects (as everything else in nuclear physics), and drastically influence the size distribution (\ref{distribution}).  This is precisely the reason why  any quantitative estimates of the size distribution are very hard to carry out.   It should be contrasted with our   total  energy budget estimates (\ref{total_power}) and the typical spectral properties (\ref{eq:Ftot}) which are   expressed in terms of the observable parameters (such as dark matter density) and   represent very solid predictions of this framework as they are  not  sensitive  to the size distribution.

 To reiterate:  it is not feasible at present time to make any considerable progress  in theoretical  estimates  of the baryon charge distribution  $dN/dB$ due to the  two  major problems  outlined above. 

Fortunately, on the observational (data analysis) side with the estimates  $dN/dW$ some  progress can be made.  In fact, we want to make few comments on the recent papers  \cite{Klimchuk:2005nx,Klimchuk:2017,Pauluhn:2006ut,Hannah:2007kw,Bingert:2012se,nanoflares} related to the nanoflares  and corresponding  statistical analysis of the observations   in the corona\footnote{I thank   anonymous referee for the relevant comments and pointing out to some important references related to the recent nanoflare observations.}.  In most  studies the term ``nanoflare"    describes   a generic event  for any impulsive energy release on a small  scale, without specifying its cause, see  review paper   \cite{Klimchuk:2005nx} and  preprint   \cite{Klimchuk:2017} 
 with references on  recent  activities in the field. In other words,  in most studies  
     the hydrodynamic consequences of impulsive heating (due to the nanoflares)  have been used without discussing their     nature. In contrast, in our  work we do not discuss the hydrodynamics, but rather we address precisely the question on the origin of the energy  responsible for coronal heating problem.
 Our total  energy budget estimate (\ref{total_power})   in the AQN model suggests that  the nuggets  might be responsible for 
 coronal heating  as the corresponding estimate  is  consistent with the observed luminosity (\ref{estimate}).
 \exclude{ Furthermore, our   interpretation  of the nanoflares 
 as the annihilation events of the AQNs  with the plasma as advocated in previous subsections  \ref{nanoflares-history}, \ref{nanoflares-nuggets}
 is also consistent with recent observations and statistical analysis. 
 }

With these preliminary remarks  it is fair to say  that there is some  agreement between different groups  \cite{Klimchuk:2005nx,Klimchuk:2017,Pauluhn:2006ut,Hannah:2007kw,Bingert:2012se,nanoflares} that nanoflares may play a dominate role in heating of solar corona.  
For example, for the typical parameters assumed in  \cite{Pauluhn:2006ut} the authors conclude that  in order to reproduce the observations the nanoflares must have typical  energies $\la W\ra$  and the frequencies of appearance $ {dN}/{dt}$ as follows  \cite{Pauluhn:2006ut}:
\be
\label{average}
\la W\ra \simeq 10^{23}~ {\rm erg} , ~~~~~~  \la \frac{dN}{dt}\ra \simeq (10^3-10^4)~ {\rm s}^{-1}.
\ee 
 The product $\la W\ra\cdot \la \frac{dN}{dt}\ra$   obviously shows that (\ref{average}) 
  is perfectly consistent with previous analysis  \cite{Benz-2000,Benz-2001, Kraev-2001,Benz-2002, Benz-2003}  overviewed   in subsection \ref{nanoflares-history} and with our estimate (\ref{total_power}) based on the AQN dark matter model. At the same time, there are some  disagreements between different groups  on spectral properties $dN/dW$ of the flares expressed in terms of power-law index $\alpha$ as defined in  (\ref{distribution}). 
 
 In particular, the authors of ref. \cite{Pauluhn:2006ut} claim that the best fit to the data is achieved with $\alpha\simeq 2.5$ with typical energies (\ref{average}), while numerous attempts to reproduce the data with $\alpha < 2$ were unsuccessful.  It should be contrasted with another analysis \cite{Bingert:2012se} which suggests that $\alpha\simeq 1.2$ for events below $W \leq 10^{24}$ erg,   and $\alpha\simeq 2.5$ for events above $W \geq 10^{24}$ erg.  Analysis \cite{Bingert:2012se}  also suggests that the change of the scaling (the position of the knee) occurs at energies close to $\la W\ra \simeq 10^{24}~ {\rm erg}$, which roughly coincides with the maximum of the energy distribution, see Fig.7 in   \cite{Bingert:2012se}. 
 
One should also remark on a different analysis \cite{Hannah:2007kw} which includes RHESSI data with the energies range from $10^{26}~ {\rm erg}$
to $10^{30}~ {\rm erg}$ with the median being about $\la W\ra \simeq 10^{28}~ {\rm erg}$. It has been noted in \cite{Hannah:2007kw} that it is conceivable 
that the distribution of all flares follows a single power-law with $\alpha\simeq 2$, which might suggest a common origin for all flares, see Fig. 18 in \cite{Hannah:2007kw}.  Of course, the comparison between different components of the energy distribution is a  highly  nontrivial procedure as it includes comparison of the data produced by the different instruments with specific instrumental effects. Furthermore,   different components of the energy distribution covers different phases of  the solar cycle. 

 Our   comment on the proposal of ref.  \cite{Hannah:2007kw}  (that all flares follow a single power-law with $\alpha\simeq 2$) can be formulated as follows. It still requires some  additional arguments and analysis to see that  the gaps between different energy components  presented on Fig. 18 in \cite{Hannah:2007kw} are indeed the instrumental, and not the physical  effects.   Furthermore, the proposal of ref.  \cite{Hannah:2007kw} is not easy to reconcile with the comment  made in \cite{Benz-2003} that ``the occurrence rates of nano flares and micro flares are so different that they cannot originate from the same population". This proposal is also difficult to reconcile with results  of ref.\cite{Bingert:2012se} where the ``knee" in the spectrum is observed when   the power law changes from $\alpha\simeq 2.5$ for high energies to $\alpha\simeq 1.2$ for low energies. From the AQN dark matter model perspectives   the nanoflares and microflares 
 are considered to be very different types of events, with the only possible connection between them  is that the   anti-nuggets could play the role of the {\it triggers}   activating   the magnetic reconnection of {\it preexisted}  magnetic fluxes in active regions (which may generate the larger flares), as remarked in item 3 in Section \ref{speculations}. 
 
   We conclude this section with the following remarks.  Our proposal (that the observed nanoflares can be interpreted as the annihilation   events of the anti-nuggets) is consistent with available observations and data analysis, including the independent constraints on the axion mass $m_a$ and the baryon charge $B$ as  mentioned in Section \ref{nanoflares-nuggets}. While a theoretical estimate for the  baryon charge distribution $dN/dB$ defined by (\ref{distribution})    is not feasible at the present time, some  further progress  in modelling of the $dN/dW$ distribution (which is proportional to $dN/dB$ within the AQN model) is a achievable, in principle,   as it   is based on conventional technical tools such as magnetohydrodynamical   simulations and data analysis,  
   as discussed in Section \ref{nanoflares-development}. Furthermore, the future advances in the axion search experiments, see recent reviews   \cite{vanBibber:2006rb, Asztalos:2006kz,Sikivie:2008,Raffelt:2006cw,Sikivie:2009fv,Rosenberg:2015kxa,Graham:2015ouw,Ringwald2016},    will further constraint the axion mass $m_a$. These constraints may narrow the window for the baryon charges $B$ of the nuggets within the framework of the AQN dark matter model.  Consequently, the narrowing window for allowed  $B$ can be  translated into the constraints on  the possible energy window   for  the nanoflare events as discussed in Section \ref{nanoflares-nuggets}.

 \section{Sun as an  ideal place to study  AQN dark matter model  }\label{sun}
 In this section we want to argue that the solar environment has a number of special features which make the Sun to be  a perfect place to study the AQN  dark matter model overviewed in Section \ref{sec:QNDM}.

  In the  previous studies   \cite{Oaknin:2004mn, Zhitnitsky:2006tu,Forbes:2006ba, Lawson:2007kp,
Forbes:2008uf,Forbes:2009wg,Lawson:2012zu} it has been argued that the AQN dark matter  model   is consistent with all presently available observations of the  galactic radiation. Furthermore, there is a number of frequency bands where some excess of emission was observed, and this model may explain some portion, or even entire excess of the observed radiation in these frequency bands.  At the same time, 
nearby galaxies are essentially not sensitive to this type of dark matter  as argued in \cite{Lawson:2015xsq,Lawson:2016mpu}. 
It has been also previously argued \cite{Lawson:2010uz,Lawson:2012vk} that the  large scale cosmic ray detectors are capable of observing quark nuggets passing through the earth's atmosphere,    see also short review paper \cite{Lawson:2013bya}. 

In contrast with those previous studies of the nuggets passing through the earth's atmosphere, our present proposal is to use the solar corona background in order to  investigate  the AQN dark matter model. This proposal   has many  advantages    in comparison with the ground based  searches  on Earth due to  the following main  reasons.  The earth's atmosphere is much smaller in height  than the solar atmosphere. Furthermore, the solar   parameter $ \gamma_{\odot} $  as defined by eq. (\ref{capture}) is numerically large.
This implies that a large portion of the nuggets which are not on head on collision trajectories nevertheless will be  captured by the sun. 
It also implies that a typical length of the nugget's trajectory is of order of $ R_{\odot}$, i.e. $\sim 10^6$ km. 

It should be contrasted with collision with the Earth when $\gamma_{\oplus} \ll 1$, and therefore only the nuggets with  head on trajectories  will be captured by the Earth. A typical length  of the nugget's  trajectory in this case of head on collisions   is of order of  height of the earth's atmosphere $\sim 10^2$ km, rather than  $R_{\oplus}$.

Another important distinct feature is that  the solar atmosphere   is the mixture of  the light elements such as hydrogen and helium which can be easily annihilated by the  anti-nuggets.
It should be contrasted with earth's atmosphere  which is a mixture of relatively heavy elements such as nitrogen and oxygen which are most likely to be reflected by the anti-nuggets rather than annihilated by them. 
Therefore, in spite  of the  fact that the column density of the earth's atmosphere is much higher than the column density of the solar atmosphere, the number of the annihilation events is expected to be much higher for the Sun in comparison with the Earth as a result of these two distinct features. 

Furthermore, the solar atmosphere is highly ionized system,  in contrast with the earth's atmosphere, and therefore, the  annihilation  processes are  much more efficient  in the solar environment in comparison with  the Earth environment due to large number of  protons from plasma accompanying the anti-nuggets as a result of  the Debye screening. This effect drastically increases the effective  radius  $R_{\rm eff}$ of the interaction of the anti-nuggets as described  in Section \ref{AQN}. 

These few  distinct features   lead to the profound observational consequences: an  anti-nugget entering the earth's atmosphere will be loosing only  a tiny  portion of its baryon charge before hitting the ground and entering the earth's dense underground regions,  after  which   the direct observable   consequences are very hard to recover.

 It should be contrasted with an anti-nugget entering the solar corona in which case the anti-nugget may   completely loose  its baryon charge even before the entering the dense regions of the photosphere as argued in section \ref{AQN}.  As a result of these  differences,    most of the energy  in the Earth's case will be deposited in the  deep earth's underground  regions while the major portion of the   energy in the solar's case will be mostly deposited  in  the  corona/ chromosphere/ transition   regions   allowing  the direct observations in form of the EUV and soft x-ray emissions  by EIT and similar instruments. 
 This makes the solar atmosphere  to become an   ideal environment to study the AQN dark matter   model.

 \section{  Concluding remarks and  Wild speculations.}\label{speculations}
 The main claim of the present work is that the observed  EUV and soft x-rays in the Corona might be originated from the AQNs proposed  in \cite{Zhitnitsky:2002qa} for completely different purposes. The    total solar  intensity (\ref{total_power}), the spectrum and estimated  temperature (\ref{eq:Ftot}) are fully consistent  with this proposal\footnote{The corresponding estimates were based on the assumption   explicitly formulated   in the first paragraph  in section \ref{interpretation}.}.  
 
 We also conjectured in   in section \ref{nanoflares} that the events of annihilations of the AQNs  can be interpreted as the  previously studied nanoflare events. 
  This conjecture on identification of these  two (naively different) entities  is supported by a number of observations listed in section \ref{nanoflares-history}
as   interpreted in section \ref{nanoflares-nuggets} in terms of the annihilation events of the anti-nuggets in the  corona /TR/ chromosphere. 

We   list below 
   a number of observed (but not understood)   phenomena  in many systems, which   could  be related to the dark matter AQNs studied in the present  work. The corresponding  list, which is obviously far from complete,  in particular, includes: 

1. There are some unexplained events, such as the Tunguska-like events when no fragments or chemical traces have ever been recovered. We speculate that such unusual events  might be the result   of the  collision of the anti-nugget with  Earth, as it was previously  mentioned in the original  paper \cite{Zhitnitsky:2002qa}.

2. There are some unexplained seismic-like events, similar to the one discussed in  \cite{Anderson:2002qx}. Such events   could  also be related to the dark matter nuggets advocated in this work. 

3. The conventional viewpoint is that the solar flares  occur as a result of  the magnetic reconnections. There are many known controversial elements in implementation of this idea which   shall not be discussed here. The only original comment we would like to make here is that the quark anti-nuggets could play the role of the {\it triggers}   activating   the magnetic reconnection of preexisted  fluxes.  This is because the nuggets locally  deposit the energy which   may produce a sufficiently strong  local disturbance initiating the flares.  It may also    speed   up the magnetic reconnection which is known to be a crucial controversial element in corresponding computations, see textbook \cite{choudhury} for review. 

4. There  is a number of studies  on the galactic scales  suggesting that   the  DM, in fact,    couples to the luminous  matter. 
 In particular, in ref. \cite{Salucci} it has been argued that the observed correlations between the dark and luminous components 
 is very hard to explain in a conventional dark matter scenario. Our original comment here is that the AQN dark matter model is a natural candidate which is, in principle,  capable to   resolve these puzzles as the nuggets are actually made of strongly interacting   quarks and gluons. Furthermore, both components of matter (visible and dark) have the same origin, have been formed at the same QCD transition, and both proportional to the same  fundamental dimensional  
 parameter $\Lambda_{\rm QCD}$ as reviewed in Section \ref{sec:QNDM}.
 
 5. A very different study of a number of  correlations in  the Sun and its planets strongly suggest a presence  of ``invisible matter"  \cite{Zioutas}. It would be very interesting to see if  the AQNs are capable to play the role of the  ``invisible matter" postulated in \cite{Zioutas}.
 
We conclude this work with following proposals   for the futures studies  which may further support (or rule out) this mechanism  when  the energy supply to the  Corona   is provided by the  AQNs as determined by the dark matter density $\rho_{\rm DM} \sim 0.3~ {\rm GeV cm^{-3}}$ according to  estimate   (\ref{total_power}).  This mechanism  automatically, without adjusting of any parameters,    generates the  observed  fraction $\sim 10^{-6}$   of the total solar luminosity emitted  in  the form of the  EUV and soft  x-rays.  
 
$\bullet$  First of all, a similar EUV and soft  x-ray radiation discussed in the present work observed in the Sun must be present 
 in many similar  stars, though the intensity and the spectral properties are highly sensitive to the specific features   of stars and their positions in the galaxy.   Such a radiation indeed  has been observed in many systems,  see \cite{testa:2015} for review. A detail studies of  the EUV and x-ray radiation (especially in quiet stellar's regions) in different types of stars is highly desirable 
 as it may shed some light on understanding of different    types of energy sources responsible for the corona heating  for variety of stars. The intensity and spectral properties of the radiation must obviously depend on   internal structure of a  star under study,  as well as the  outer dark matter density $\rho_{\rm DM}(r)$ which itself strongly depends on  position of the  star with respect to the galactic center. 
 
 In particular, if the future observations   suggest that 
  the EUV and x-ray radiation from stars is not sensitive to surrounding   dark matter density $\rho_{\rm DM}(r)$ it would unambiguously rule out our proposal as the number of annihilation events (and therefore the intensity of the EUV radiation) is directly proportional to the dark matter density in vicinity of a  specific star  according to the basic formula  (\ref{total_power}).

$\bullet$ Secondly,   the observation of the nanoflares
 with SoHO/EIT  has a specific instrumental threshold $\sim 3\cdot 10^{24}$ erg. A very modest improvements in resolution (on the level of factor 3 or so) should  result in corresponding increase of the rate of the observed  nanoflares, which we identify with annihilation events of anti-nuggets with solar material as described in section \ref{nanoflares}.
 The corresponding measurements would provide the information about the nugget's  size distribution ({\ref{distribution}), and may resolve some discrepancies between different group modelling and analyzing the data as mentioned in  section \ref{nanoflares-development}.
 It is very important to  perform  the corresponding analysis by separating the nanoflares in quiet regions from active regions to avoid any mis-identification of the nanoflares with unrelated  micro-events which  receive their energy as a result of the alternative sources such as the   magnetic reconnections  in active regions. 
 
 In particular, if the future observations show   that there is a sharp cutoff (or other drastic changes) in the nanoflare distribution somewhere  in the range $(10^{22}-10^{24}) $ erg it would unambiguously rule out the proposal because the nugget's baryon charge $B$ must be continuously extended to sufficiently low $B\sim 10^{24}$
 as the average baryon charge of the  the AQNs  is estimated to be  in the range  $\la B\ra \sim 10^{25}$ as mentioned at the  end of section \ref{sec:QNDM}.

\section*{Acknowledgements} 
 I am thankful to Konstantin Zioutas for inviting me to the Patras workshop 2017 (Thessaloniki, May 2017) where this project was started.  
I am also grateful to him for  our collaboration during the initial stage of this project. I am also indebted to Konstantin for teaching me about the solar system and its mysterious behaviour. I am also thankful to Kyle Lawson for useful comments. 
 This research was supported in part by the Natural Sciences and Engineering 
Research Council of Canada.

\appendix

 \section{Few technical characteristics of the   anti-nuggets at $T\neq 0$}\label{ionization}
 The main goal of this Appendix  is to provide few simple qualitative estimates which have been used in the main text.  
 
 First, we 
 estimate the electrical charge of the anti nuggets when they enter the solar corona. The basic idea of the estimate is as follows. The total neutrality of the nuggets in the model is supported by electrosphere made of leptons (positrons for the anti-nuggets). For non-zero intrinsic nugget's temperature $T\neq 0$   a small portion of the loose positrons will be stripped off from the anti-nuggets. As a result the anti-nuggets will  esquire a non vanishing negative electric charge $Q$. To estimate this charge $Q$ one can use the electro-sphere density profile function $n(r)$ by removing the contribution of the region of loosely bounded  positrons with  low momentum $p^2\leq 2 m_e T$. 
 
 The corresponding computations can be carried out using the electro-sphere density profile function computed in 
  \cite{Forbes:2009wg} within  the Thomas-Fermi approximation\footnote{The profile function computed in 
  \cite{Forbes:2009wg}  describes well the interpolation between the non-relativistic and  relativistic regimes. It was important for relating the intensities of the galactic  of 511 keV line with  $ (1-20)$ MeV diffuse photons which resulted from annihilation of visible electrons with   non-relativistic  positrons (511 keV line) and relativistic positrons (in $ (1-20)$ MeV range) from the electro-sphere.}. For the present work it is sufficient to use the non-relativistic approximate expression   for the profile function  $n(r)$  given in  \cite{Forbes:2008uf}. The corresponding computation leads to the following estimate  for $Q$: 
  \be
  \label{Q}
Q\simeq 4\pi R^2 \int^{\infty}_{\frac{1}{\sqrt{2 m_e T}}}  n(z)dz\sim \frac{4\pi R^2}{2\pi\alpha}\cdot \left(T\sqrt{2 m_e T}\right).~~~~~
  \ee
  For the numerical estimate we assume $R\sim 10^{-5}{\rm cm}$ and $T\simeq 100$ eV according to eq. (\ref{eq:Ftot}). In this case $Q\sim 10^{8}$ which represents very tiny portion in comparison with  the baryon charge $B\sim 10^{25}$ hidden in the anti-nugget, i.e. $(Q/B)\ll 1$. Nevertheless, the charge $Q$ plays an important role in the dynamics of the anti-nuggets as they become   subject to  the strong electric and magnetic   forces, which are ubiquitous in the Sun atmosphere.

 The second parameter which was  used in  our estimates (\ref{time}), (\ref{time1})  in the main text is the effective size $R_{\rm eff}(l)$ of a nugget when it moves in the hot 
 ionized plasma. As explained in the main text  this environment may drastically  affect   the  annihilation rate due to a   number of  many body plasma phenomena    mentioned after eq. (\ref{time}). The same scale $R_{\rm eff}(l)$ essentially determines how the generated  energy (due to the annihilation processes) is transferred to the surrounding plasma.   One can treat the parameter $R_{\rm eff}(l)$ as a typical scale which counts the number of particles from plasma $\sim n_{\rm sun}(l)R_{\rm eff}^3(l)$ which effectively participate in the processes of annihilation and   energy transfer from the nugget to surrounding plasma.
 
 The simplest and very rough way to estimate the corresponding effective size $R_{\rm eff}$ is to approximate  an effective Coulomb cross section by assuming that   a typical  momentum transfer
 is order of the temperature of the surrounding plasma,  $|q|\sim T$, i.e.
 \be
 \label{momentum}
 \pi R_{\rm eff}^2\sim \frac{Q^2\alpha^2}{q^2}\sim \frac{Q^2\alpha^2}{T^2}.
 \ee
 Using numerical estimates for $Q$ from (\ref{Q}) and for temperature $T\simeq 10^6 {\rm K}$ we arrive to the following  estimate for $R_{\rm eff}$
 which effectively determines the  size of the  system
 \be
 \label{R_eff}
  R_{\rm eff}\sim 0.1~ {\rm cm}~~~  {\rm for} ~~Q\sim 10^8~~ {\rm and}~~  T\simeq 10^6 {\rm K}.
 \ee
  Precisely this estimate has been used in formula (\ref{time1}) in the main text. 
  
  One should emphasize that $R_{\rm eff}$ estimated above is drastically different from  our previous  studies  (when the relevant  size coincides with the radius  of the nugget $R$) reviewed in section \ref{sec:QNDM} due to two reasons.  First, the  anti-nuggets are the charged objects according to (\ref{Q}). Secondly, the charged anti-nuggets 
 are propagating in  ionized plasma which strongly affects the annihilation rate  as well as the energy transfer rate, as mentioned above. These features make the solar atmosphere to become an   unique place to study this type of dark matter, as discussed in section \ref{sun}.
  
 \exclude{
 The enhancement parameter $\eta(T)$ introduced above in eq.(\ref{eq:Fann1})  is expected to be very large in these circumstances as a result of these continuous   processes.  These effects which occur in highly ionized plasma of the  solar Corona are drastically different from
 analysis of  the annihilation pattern which occurs in  the galactic environment and in the earth's atmosphere when the anti-nuggets can be treated as the  macroscopically  large and electrically neutral  objects as argued  in Section \ref{speculations}. 
}


\end{document}